# Double folding analysis of $^3$He elastic and inelastic scattering to low lying states on $^{90}$Zr, $^{116}$Sn and $^{208}$Pb at 270 MeV

**Marwa N. El-Hammamy**

*Physics Department, Faculty of Science, Damanhur University, Egypt*

**Abstract**

The experimental data on elastic and inelastic scattering of 270 MeV $^3$He particles to several low lying states in $^{90}$Zr, $^{116}$Sn and $^{208}$Pb are analyzed within the double folding model (DFM). Fermi density distribution (FDD) of target nuclei is used to obtain real potentials with different powers. DF results are introduced into *modified DWUCK4 code to* calculate the elastic and inelastic scattering cross sections. Two choices of potentials form factors; Woods Saxon (WS) and Woods Saxon Squared (WS$^2$) for real potential are used, while the imaginary part is taken as phenomenological Woods Saxon (PWS) and phenomenological Woods Saxon Squared (PWS$^2$). This comparison provides information about the similarities and differences of the models used in calculations.

## 1. Introduction

When we bombard a nucleus with a nucleon or with light ions like d, $^3$He, α-particles, etc., various nuclear phenomena occur, including elastic scattering, inelastic scattering, nucleon transfer reactions and projectile fragmentation, depending on the projectile species and the bombarding energy. The simplest among these phenomena is the elastic scattering. Elastic scattering can provide valuable information about the interaction potential between two colliding nuclei [1]. Inelastic scattering of $^3$He particles belongs to useful methods for investigation of excited states of nuclei. According to the importance of these reactions, the analysis of experimental data on elastic and inelastic scattering of $^3$He from different targets is needed [2-4]. The main problem of investigating the light heavy ion reactions by using nuclear reaction models is to determine the most suitable potential form to explain the experimental data. Optical (OM) and Folding models (FM) are examples of simplified models that exist for studying light heavy ion reactions [1,5-8]. The OM has a potential including the real and the imaginary potentials. The real potential describes the elastic scattering of the reaction. The imaginary potential expresses the loss of flux into non elastic channels. The real and imaginary potentials can be determined with either the phenomenological or the microscopic model.

In the microscopic model, while the imaginary potential is taken PWS or PWS$^2$ type potential, the real potential can be defined using DFM. In DFM, the density distributions (DD) of both projectile and target nuclei are used. Therefore, DD used in double folding calculations is very important in examining of nuclear reactions.

The purpose of this work is to analyze angular distributions of the elastic and inelastic scattering of $^3$He with an energy of 270 MeV leading to the excitation of $^{90}$Zr levels, 2.18 MeV ($2^+$), 2.75 MeV ($3^-$),$^{116}$Sn levels 1.29 MeV ($2^+$), 2.27 MeV ($3^-$), and $^{208}$Pb levels 4.09 MeV ($2^+$), 2.61 MeV ($3^-$) MeV [2] in the framework of DFM by using FDD of target nuclei and WS potential forms with different powers (n =1 or 2) as case one and case two. The importance of inelastic scattering analysis to low lying states is to test the strength of these states within DFM.

The method employed here and discussions are given in Section 2, and the conclusions are presented in Section 3.

## 2. Analysis and discussion

To study the *elastic scattering* for the reactions of $^3$He - particles with $^{90}$Zr, $^{116}$Sn and $^{208}$Pb, the program code *DFPOT* [9] has been used. V(r) is the DF potential carried out by introducing the effective nucleon-nucleon (NN) interaction over the ground state DD of the two colliding nuclei. It is evaluated from the expression

$$V(r) = \int\int \rho_1(r_1)\rho_2(r_2)V_{NN}(s)dr_1dr_2 . \quad (1)$$

$\rho_1(r_1)$ and $\rho_2(r_2)$ are the nuclear matter density of the two colliding nuclei, and $V_{NN}(s)$ is the effective NN interaction potential ( $s = r - r_2 + r_1$ ). $V_{NN}(s)$ is taken to be a standard Reid- M3Y interaction [10] in the form,

$$V_{NN}(r) = 7999.0\frac{e^{-4.0r}}{4.0r} - 2134.0\frac{e^{-2.5r}}{2.5r} + J_{00}(E)\delta(r) . \quad (2)$$

The first and second terms represent the direct part and the third term represents the exchange part of the interaction potential. It plays an important role in reproducing the experimental results for elastic and inelastic scattering [11,12]. The exchange part can be written to a good approximation in the form [10]

$$J_{00}(E) = -276(1 - 0.005\frac{E}{A}) , \quad (3)$$

where E is the energy in the center of mass system and A is the mass number of the projectile.

In our calculations, the nuclear matter DD of $^3$He nucleus has the Gaussian form

$$\rho = \rho_\circ \exp(-\alpha r^2) , \quad (4)$$

where $\alpha = 0.5505$ $fm^{-2}$, $\rho_\circ = 0.2201$ $fm^{-3}$ [13], and for $^{90}$Zr [14], $^{116}$Sn and $^{208}$Pb [15] the following FDD form is used

$$\rho = \rho_\circ \left[1 + \exp\left(\frac{r-R}{a}\right)\right]^{-n} , \quad (n = 1 \text{ or } 2) \quad (5)$$

The total potential must comprise both the real part and the imaginary part, the latter is being responsible for the absorption of the incident particle in the inelastic channels.

$$U(r) = N_r V(r) + iW(r) \quad (6)$$

Since the M3Y interaction is real, the folding calculation gives the real part of the potential. In the model used here, the volume real part has the folded form with normalization factors $N_r$. We have chosen this form to be WS shape, while the imaginary part is taken as PWS. The resulted folded form factors, in addition with PWS potentials parameters in each case (n=1 or 2) are introduced into the modified program code *DWUCK4* [16] to compute the differential scattering cross section, in which an additional form factor form $(WS)^2$ is added. The calculations for elastic scattering were calculated by DWUCK4. *Firstly*, we used (WS) for real and (PWS) for imaginary parts of the potential as case one. *Secondly,* in case two, we used $(WS)^2$ for real and $(PWS)^2$ for imaginary parts of the potential. Thus, the real potentials are represented by

$$V(r) = -V_\circ \left[1 + \exp(\frac{r - R_v}{a_v})\right]^{-n},$$

while the imaginary potentials are represented by

$$W(r) = -W_\circ \left[1 + \exp(\frac{r - R_w}{a_w})\right]^{-n} \quad (7)$$

in which $R_{v,w} = r_{v,w}(A_T^{1/3} + A_P^{1/3})$ and $n = 1\ or\ 2$. $V(r)$ is the DF potential of equation (1) and $N_r, W_o, r_w, a_w,$ are variable parameters. Comparisons are shown in figure (1) between the present calculations and experimental data.

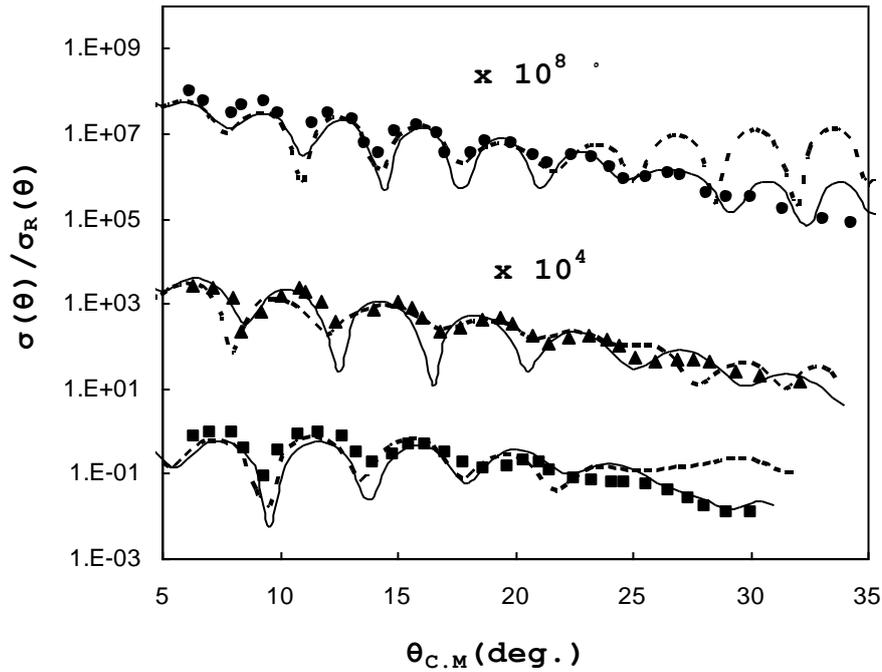

**Figure (1).** Angular distributions of $^3$He elastically scattered on $^{90}$Zr, $^{116}$Sn and $^{208}$Pb. The theoretical cross section obtained with DF model is represented by *dotted lines* for case one and *solid lines* for case two. Experimental points are denoted by *black symbols*, ■ for $^{90}$Zr, ▲ for $^{116}$Sn and ● for $^{208}$Pb.

The variable parameters of the two cases and DF-equivalent potential parameters ($V_o, r_v, a_v$) are listed in table (1).In order to estimate the quality of the fit, one can calculate a relative error

$$\chi_R^2 = \frac{1}{N} \sum_{i=1}^{i=N} \left[ \frac{(\sigma^{calc.}(\theta_i) - \sigma^{exp.}(\theta_i))}{(\sigma^{calc.}(\theta_i) + \sigma^{exp}(\theta_i))} \right]^2 \quad , \tag{8}$$

where N is the number of data points and $\sigma^{calc.}(\theta_i)$ is the i$^{th}$ calculated scattering cross section and $\sigma^{exp.}(\theta_i)$ is the corresponding experimental scattering cross section.

*Table (1):* The real and imaginary potentials parameters of $^3$He elastic scattering on different nuclei

| Reaction | n | $V_o$ (MeV) | $R_v$ (fm) | $a_v$ (fm) | $N_r$ | $W_0$ (MeV) | $r_w$ (fm) | $a_w$ (fm) | $\chi_R^2$ |
|---|---|---|---|---|---|---|---|---|---|
| $^3$He+ $^{90}$Zr | 1 | 143 | 4.292 | 1.228 | 0.78 | 20 | 1.20 | 0.24 | 0.19 |
|  | 2 | 209 | 5.011 | 1.528 | 1 | 29 | 1.20 | 0.54 | 0.09 |
| $^3$He+ $^{116}$Sn | 1 | 143.4 | 4.795 | 1.238 | 0.69 | 20 | 1.14 | 1.0 | 0.11 |
|  | 2 | 205.3 | 5.517 | 1.529 | 1 | 52 | 1.14 | 1.19 | 0.10 |
| $^3$He+ $^{208}$Pb | 1 | 143.8 | 6.136 | 1.222 | 0.69 | 20 | 1.18 | 0.24 | 0.31 |
|  | 2 | 192.3 | 6.861 | 1.486 | 1 | 35 | 1.18 | 0.66 | 0.15 |

For first case, the agreement of the theoretical angular distribution with the experimental one is excellent at forward angles $\theta_{c.m} < 22°$, then discrepancy appeared in larger angular regions. Thus, these results should be improved with another theoretical approach. Therefore *secondly in case two*, we increased the power of the potential form to be squared as it was successful in many other analyses within OM [17,18]. According to an increase of real normalization factor ($N_r = 1$) values by increasing power (n), the results are better than in case one, with less relative error $\chi_R^2$ values. This is an important point in studying the interaction of $^3$He. Because, if it is investigated the interaction of $^3$He with different target nuclei within framework of DFM, it is mostly needed normalization to obtain a satisfied agreement results with the experimental data.

Thus, we used the case two potential parameters to be used in inelastic scattering analysis. The difficulty found in fitting elastic scattering cross sections is reflected in the inelastic predictions and indicated a deficiency in the present potential form.

The analysis of the *inelastic scattering* of the $^3$He particles has been performed and the comparison of theoretical calculations and the experimental data has been presented in figure (2).

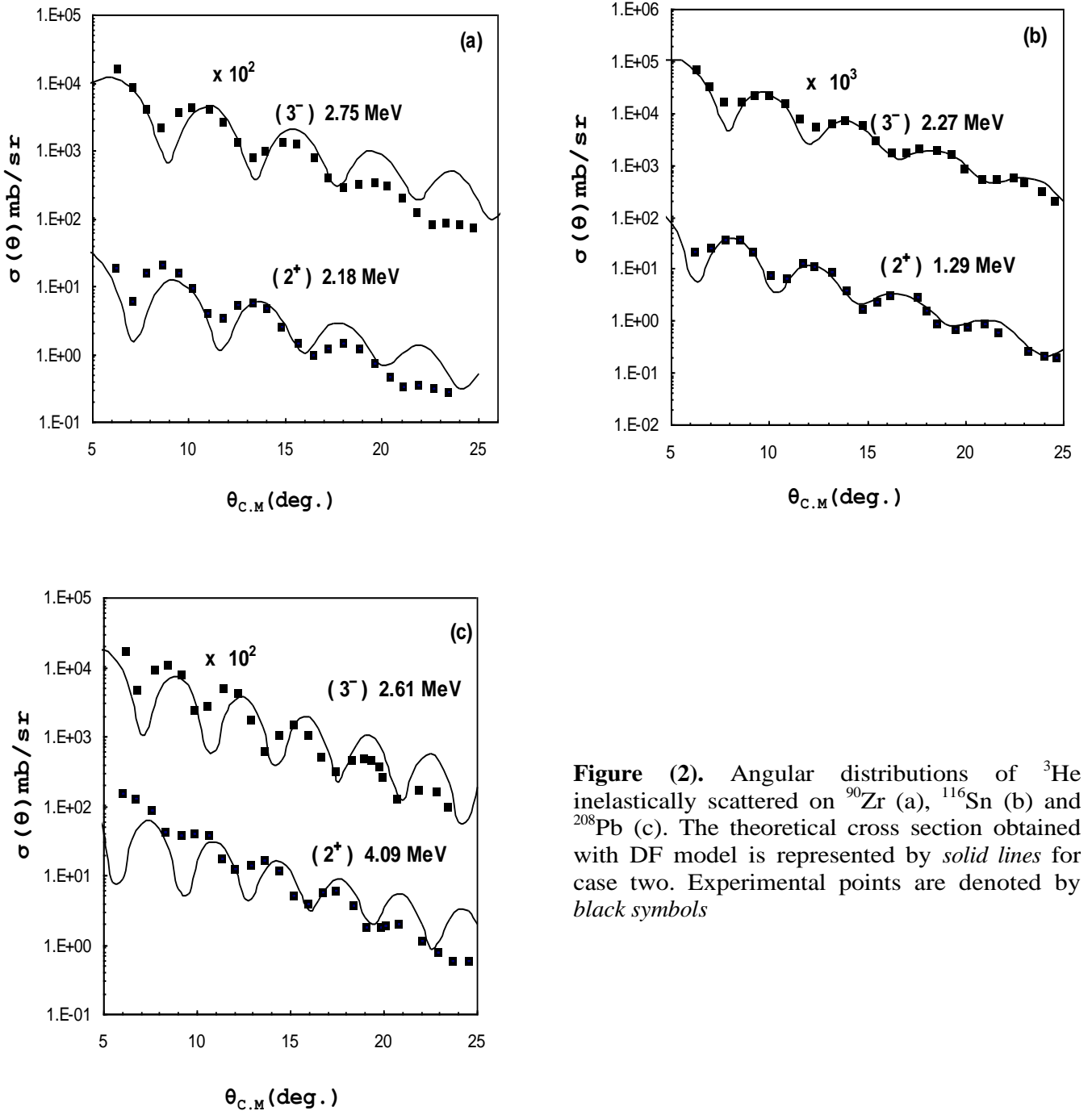

**Figure (2).** Angular distributions of $^3$He inelastically scattered on $^{90}$Zr (a), $^{116}$Sn (b) and $^{208}$Pb (c). The theoretical cross section obtained with DF model is represented by *solid lines* for case two. Experimental points are denoted by *black symbols*

In case of $^{90}$Zr and $^{208}$Pb ($3^-$) state, there is an overestimation in forward regions and poor agreement in case $^{208}$Pb ($2^+$) state. The potentials for elastic scattering analysis are subsequently used to calculate the inelastic scattering cross sections in the modified DWUCK4. The inelastic potentials are calculated according to the following form

$$U^\lambda(r) = N_{rr} V^\lambda(r) + i W^\lambda(r), \qquad (9)$$

where λ is the multi-polarity [19]. $V^\lambda(r)$ is the real folded (transition) inelastic potential multiplied by normalization factor $N_{rr}$ and $W^\lambda(r)$ is an imaginary deformed PWS potential.

The calculated real folded inelastic potential normalization factor $N_{rr}$ as well as the corresponding values of $\chi_R^2$ are shown in table (2).

**Table (2):** *Real normalization factor and $\chi_R^2$ values from best fit to inelastic scattering data for different levels of $^{90}Zr$, $^{116}Sn$ and $^{208}Pb$*

| Reaction | n | Level | $N_{rr}$ | $\chi_R^2$ |
|---|---|---|---|---|
| $^3He + {}^{90}Zr$ | 2 | 2.18($2^+$) | 0.13 | 0.14 |
|  |  | 2.75($3^-$) | 1.40 | 0.14 |
| $^3He + {}^{116}Sn$ | 2 | 1.29($2^+$) | 0.17 | 0.03 |
|  |  | 2.27($3^-$) | 0.84 | 0.03 |
| $^3He + {}^{208}Pb$ | 2 | 4.09($2^+$) | 0.23 | 0.19 |
|  |  | 2.61($3^-$) | 1.44 | 0.16 |

## 3. Conclusion

Although there are many detailed analyses concerning the elastic and inelastic scattering angular distributions of these investigated systems studied in OM with various potential forms, just a few of them make an effort to evolve a systematization for the folding potential parameters. So, we have reanalyzed elastic and inelastic scattering of $^3$He - particles with $^{90}$Zr, $^{116}$Sn and $^{208}$Pb at 270 MeV with minimal 4-parameter nuclear potential sets having (WS$^2$+iPWS$^2$) and (WS+iPWS) forms. When the real potential parameters are used with different normalization factors given in this work, FM analyses with these two folded potential sets have provided different results. Calculations with the squared potential forms can reproduce the experimental elastic angular distributions in a good agreement, especially in case $^{116}$Sn. The difficulty found in fitting elastic scattering cross sections is reflected in the inelastic predictions in case $^{90}$Zr and $^{208}$Pb and indicated a deficiency in the present potential form.

However, this approach has shown that an increase of power (n) from 1 to 2 is accompanied by an increase of real normalization factor ($N_r = 1$ for all cases) values, i.e. it doesn't need normalization to fit the data. The similarities and differences between the two cases used in our analysis are pleasantly visible in this comparison. Generally on the basis of these results, we conclude that that WS$^2$ form is more suitable than the PWS form for real potential.

## Acknowledgment

Author thanks the referee for valuable discussion and comments. The author would like to thank the researcher Ahmed Fouad (Faculty of education, physics and chemistry department, Alexandria University, Egypt) for providing an additional form factor form (WS)$^2$ in DWUCK4 program .

**References**

1. G. R. Satchler, Direct Nuclear Reactions, Clarendon Press, Oxford, (1983) 392.
2. P.P.Singh, Q.Li, P. Schwandt, W.W.Jacobs, M.Saber, E.J.Stephenson, A.Saxen and S.Kailas, Pramana J.Phys.27(6),(1986) 747.
3. T.Yamagata, H.Utsunomiya, M.Tanaka, S.Nakayama, N.Koori, A.Tamii, Y.Fujita, K.Katori, MInoue, M.Fujiwara and H.Ogata, Nucl. Phys.A 589 (1995) 425
4. Y.Sakuragi and M.Katsuma , Nuclear Instruments and Methods in Physics Research A 402 (1998) 347.
5. G. R. Satchler, Introduction to Nuclear Reactions, Mc Millan Press Ltd.,London (1980) 153.
6. K.S.Krane, Introductory nuclear Physics, John Wiley and Sons, New York (1988) 396.
7. M.E. Brandan, G.R. Satchler, Phys. Rep. 285, (1997) 143.
8. M. N. El-Hammamy, A. Attia, F. A. El-Akkad, A. M. Abdel-Moneim, Chinese Physics C 38, 3 (2014) 034102.
9. J. Cook, Comput. Phys. Comm. 25,(1982)125.
10. G.R. Satchler, W.G. Love, Phys. Rep. 55, (1979) 183.
11. M.El-AzabFarid, Z.M.M.Mahmoud,G.S.Hassan, Nucl.Phys.A691(2001)671 ;
12. M.El Azab Farid , Z.M.M.Mahmoud, G.S.Hassan, Phys.Rev.C64(2001)014310.
13. F.S.Chwieroth, Y.C.Tang, D.R. Thompson, Phys. Rev. C9(1974) 56.
14. M.El-Azab Farid and M.A.Hassanain, Nucl.Phys. A678,39(2000).
15. C.W.DeJager ,H.DeVries, , C.DeVries, Atomic Data and Nuclear Data Tables 14 (1974) 479.
16. P.D. Kunz, University of Colorado (unpublished).
17. M.E. Kürkçüoğlu., H.Aytekin, I. Boztosun, Modern Physics Letters A, 21 (29) 2217.
18. A. Budzanowski, C. Alderliesten, J. Bojowald, W. Oelert, P. Turek, H. Dabrowski and S. Wiktor, Proceedings of the Karlsruhe International Discussion Meeting (1979) 219.
19. D.T.Khoa ,G.R.Satchler, Nucl. Phys.A 668 (2000) 3-41.